\documentclass[10pt,english]{IEEEtran}
\usepackage{mathptmx}
\usepackage[T1]{fontenc}
\usepackage[latin9]{inputenc}
\usepackage{array}
\usepackage{float}
\usepackage{textcomp}
\usepackage{mathtools}
\usepackage{multirow}
\usepackage{amsmath}
\usepackage{amssymb}
\usepackage{graphicx}
\usepackage{rotating}
\usepackage{microtype}

\makeatletter

\providecommand{\tabularnewline}{\\}
\floatstyle{ruled}
\newfloat{algorithm}{tbp}{loa}
\providecommand{\algorithmname}{Algorithm}
\floatname{algorithm}{\protect\algorithmname}

\usepackage[unicode=true, bookmarks=true,bookmarksnumbered=false,bookmarksopen=false, breaklinks=false,pdfborder={0 0 0},pdfborderstyle={},backref=false,colorlinks=false]{hyperref}
\hypersetup{pdftitle={High performance volume ray casting:A branchless generalized Joseph projector},pdfauthor={Jonas Graetz}}

\usepackage{algpseudocode}
\newcommand{\algvspace}{\vspace{\parsep}}
\usepackage{setspace}
\usepackage{multicol}

\newcommand{\appropto}{\mathrel{\vcenter{
  \offinterlineskip\halign{\hfil$##$\cr
    \propto\cr\noalign{\kern2pt}\sim\cr\noalign{\kern-2pt}}}}}

\usepackage{scalerel}
\usepackage{tikz}
\usetikzlibrary{svg.path}
\definecolor{orcidlogocol}{HTML}{A6CE39}
\tikzset{
  orcidlogo/.pic={
    \fill[orcidlogocol] svg{M256,128c0,70.7-57.3,128-128,128C57.3,256,0,198.7,0,128C0,57.3,57.3,0,128,0C198.7,0,256,57.3,256,128z};
    \fill[white] svg{M86.3,186.2H70.9V79.1h15.4v48.4V186.2z}
                 svg{M108.9,79.1h41.6c39.6,0,57,28.3,57,53.6c0,27.5-21.5,53.6-56.8,53.6h-41.8V79.1z M124.3,172.4h24.5c34.9,0,42.9-26.5,42.9-39.7c0-21.5-13.7-39.7-43.7-39.7h-23.7V172.4z}
                 svg{M88.7,56.8c0,5.5-4.5,10.1-10.1,10.1c-5.6,0-10.1-4.6-10.1-10.1c0-5.6,4.5-10.1,10.1-10.1C84.2,46.7,88.7,51.3,88.7,56.8z};
  }
}

\newcommand\orcidicon[1]{\href{https://orcid.org/#1}{\mbox{\scalerel*{
\begin{tikzpicture}[yscale=-1,transform shape]
\pic{orcidlogo};
\end{tikzpicture}
}{|}}}}

\usepackage{fancyhdr}
\makeatother

\usepackage{babel}
\begin{document}
\title{High performance volume ray casting:\\
A branchless generalized Joseph projector}
\author{Jonas Graetz\orcidicon{0000-0002-4403-3686}\thanks{J. Graetz is with the Chair for X-ray Microscopy at the University
of Würzburg and the Fraunhofer nano-tomography group of the magnetic
resonance and X-ray imaging department MRB of the Fraunhofer IIS/EZRT
in Würzburg, Germany. (email: jonas.graetz@physik.uni-wuerzburg.de)}\thanks{Acknowledgements: The anonymous reviewers of the initial manuscript
are acknowledged for their constructive comments that have considerably
shaped the following presentation. R.\ Hanke and S.\  Zabler are
acknowledged for facilitating the present work. Funding is acknowledged
from the Bavarian State Ministry of Economic Affairs, Infrastructure,
Transport and Technology which supported the project group \textquotedblright Nano-CT
Systems for Material Characterization'', and by the German Federal
Minisry of Education and Research (grant 05E19AN1) supporting the
BM18 beamline at the European Synchrotron Radiation Facility.}}
\maketitle
\begin{abstract}
A concise and highly performant branchless formulation of a Joseph-type
interpolating ray-casting algorithm for the computation of X-ray projections
is presented. It efficiently utilizes the hardware resources of modern
graphics processing units at the scale of their theoretic maximum
performance reaching access rates of 600$\,$GB/s within read-and-write
memory, and is further shown to do so without compromising on image
quality. The computation of X-ray projections from discrete voxel
grids is an ubiquitous task in many problems related to volume image
processing, including tomographic reconstruction and visualization.
Although its central role has given rise to numerous publications
discussing the optimal modeling of ray-volume intersections, a unique
benchmark in this respect does not exist. Here, a 3D Shepp-Logan phantom
is used, which allows the computation of analytic reference projections
that can further serve as input to iterative reconstructions without
committing the inverse crime. The proposed algorithm (GJP) is compared
to the competing and widely adopted digital differential analyzer
(DDA), which computes exact line-box intersections. It is thereby
found to outperform the DDA on recent graphics processors in all respects:
Despite accessing twice as much memory, the GJP is still able to calculate
projections twice as fast. It further exhibits considerably less discretization
artifacts, and neither oversampling of the DDA nor a smooth interpolation
kernel within the GJP are able to improve on these results in any
respect.\thispagestyle{fancy}

\end{abstract}

\section{Introduction}

The simulation of X-ray images (or the generation of ``digitally
reconstructed radiographs'') by numeric projection of gridded volume
images represents, in the context of computed tomography, the calculation
of the forward problem within iterative solutions of the inverse problem,
i.e., the reconstruction problem. It is thus also referred to as ``forward
projection'' (as opposed to the ``backprojection'' step) and is
both one of the most essential and time consuming aspects of iterative
reconstruction techniques. Forward or volume projection therefore
takes a central role with respect to both efficiency and quality of
these algorithms

Foremost, simulated X-ray projection involves tracing rays through
volumes based on given projection geometries and numeric integration
of image data along these ray paths. Irrespective of any additional
features, the fundamental component of every X-ray imaging model
therefore is an adequate sampling and accumulation strategy for the
evaluation and integration of values from three dimensional voxel
grids. As many samples -- between $10^{8}$ and $10^{10}$ for typical
volume sizes of $500^{3}$ to $2000^{3}$ voxels -- are required
to compute 2D X-ray projections, and thousands of such projections
are required within iterative tomographic reconstruction, efficiency
of the sampling and integration process is of outmost importance.
The sampling strategy further affects the outcome of iterative reconstruction
algorithms, which are based on optimizing the similarity between simulated
and actual X-ray projections. Both aspects -- efficiency and physical
modeling -- have given rise to a considerable body of literature.
As the constraints and capabilities of computing hardware are constantly
evolving, the quest for most efficient solutions remains a timeless
task though. 

\subsection{Contributions}

A formulation of a 3D generalization of Joseph's classic interpolating
projection method is given and discussed. It is shown to feature excellent
memory access efficiency without explicitly restricting the projection
geometry nor making use of sophisticated memory layout schemes or
read-only texture memory. The contribution is twofold: On the one
hand, a concise and efficient algorithm is derived, benchmarked and
provided in an easily implementable form, ensuring its practical availability.
Likewise importantly, its qualitative eligibility with respect to
volume projection and iterative tomographic reconstruction as compared
to more complex approaches is assessed in order to establish it as
not only extremely fast, but also competitive despite its intriguing
simplicity. As no unique benchmark exists in this respect, a survey
of previous literature is given on the one hand, and selected experiments
demonstrating and comparing discretization artifacts of several approaches
are shown on the other hand. The initial manuscript has previously
been made available by the author as preprint \cite{DittmannGraetz2016},
and first results have been presented at the Fully3D conference \cite{DittmannGraetzHanke2017}.

\section{Literature Review}

Two general classes of volume projection approaches may be distinguished
upfront: those following integration paths and performing some kind
of sampling on the volume image grid, and those iterating over volume
elements and accumulating renderings of each voxel's projection onto
the detection screen based on the projection perspective and a model
of the voxels' geometry. The first approaches are referred to as ``ray
driven'', ``ray casting'' or ``ray tracing'' methods, while the
latter methods are usually termed ``voxel driven'' or ``splatting''.
The methods first of all differ in their memory access pattern: while
ray driven methods iterate over camera pixels and generally require
less efficient random read access to the volume image data, splatting
methods can sequentially iterate over the volume elements, yet instead
require a large amount of non-sequential read and write accesses to
the projection image. Intermediate approaches are resampling strategies
(e.g., the shear-warp approach \cite{LacrouteLevoy1994}) and the
more recent ``distance driven'' method \cite{deManBasu2004,Liu2017}.
The former transform the volume image such that the subsequent projection
reduces to a summation over one coordinate axis and are most similar
to ray driven methods. The latter approach aims to combine the sequential
memory access pattern of voxel driven methods with sequential write
accesses to the projection image. 

Ray driven projection has two important advantages, wherefore it will
be the method of choice here: first, it is trivially parallelizable,
as by design no concurrent write accesses to the projection screen
need to be managed. Secondly, the correct normalization of ray integrals
with respect to the associated run lengths through the volume is considerably
simpler as compared to splatting approaches. Simplicitly is a key
to efficiency, and it will be shown that highly efficient memory access
patterns are possible also with ray driven approaches.

Tracing of linear paths through grids has been studied since the advent
of raster graphics. The following review shall provide a reasonable
overview of the essential ideas that have come up in the past, with
a particular focus on the tomography context.

A central concept in the majority of fast ray casting algorithms on
regular grids is the notion of a ``driving axis'' \cite{Joseph1982,Lo1988,Fujimoto1986,EndlSommer1994,LiuZalik2008}
as already introduced in the 1960s by Bresenham in the context of
rasterized line drawing. Instead of just arbitrarily defining a number
of sampling points along the linear coordinate of an integration path,
the path will rather be traversed in unit steps of the designated
driving axis of the algorithm. The driving axis is chosen to be the
dimension along which the considered path progresses fastest, such
that the resulting non-integer step sizes along the remaining coordinate
axes are always guaranteed to not exceed the grid spacing, thereby
ensuring that no intersected pixels or voxels will be skipped in the
tracing procedure. Figure~\ref{fig:gridaligned-raycasting} gives
an illustration.

This concept of grid-aligned sampling is explicitly or implicitly
used e.g.\ by Josephs' algorithm \cite{Joseph1982} (one of the early
methods proposed for 2D iterative tomographic reconstruction), by
shear-warp resampling techniques \cite{CameronUndrill1992,LacrouteLevoy1994}
(proposed for volume visualization) or ray-driven formulations of
splatting algorithms \cite{MatejLewitt1996,MuellerYagel1996,Bippus2011}
as well as by the recent ``Distance Driven Method'' \cite{deManBasu2004,Liu2017}.
It emerges naturally from practical sampling considerations, as interpolation
can thereby be avoided along the driving axis. Prominent alternative
techniques are the much-cited algorithm by Siddon \cite{Siddon1985}
and variants thereof \cite{AmanatidesWoo1987,Jacobs1998,ZhaoReader2003,deGreef2009,Xiao2012}
(known as digital differential analyzer or DDA algorithm in the field
of computer graphics), which trace lines in irregular steps from intersection
to intersection with any of the raster planes perpendicular to the
coordinate axes. The final objective of calculating exact ray-box
intersections though can as well be achieved with driving-axis based
algorithms \cite{Lo1988,Gao2012}, although the complexity increases
in the 3D case.

In addition to basic algorithmic concepts, the assumed underlying
system model is a central aspect. Particularly prevalent is the assumption
that imaged objects can be exactly modeled by cubic voxels of homogeneous
density, and incident radiation by rectangular beam profiles of finite
extent (as opposed to the also common assumption of pencil beams,
cf.\ Siddon \cite{Siddon1985}). Much effort has been put into the
development of exact projection algorithms in this respect \cite{Lo1988,Gao2012,YaoLeszczynski2009,LongFessler2010,WuFessler2011,NguyenLee2012,Zhang2014,HaMueller2015,HaMueller2016,SampsonFessler2016},
using both ray driven and splatting approaches. When arguing that
there is no outstanding reason to assume homogeneous cubic voxels,
the complexity for an ``exact'' volume projector can be reduced
by using algorithmically more convenient voxel basis functions as
compared to the box profile. Modeling of both voxel and beam profiles
can then be merged into diffused, overlapping interpolation kernels
or projection footprints parametrized by ray-voxel distances \cite{MuellerYagel1996,HansonWecksung1985,Lewitt1992,Ziegler2006,MomeyDesbat2015}.
Other methods replace the latter distance by even more efficient approximations
\cite{Joseph1982,deManBasu2004,Sunnegardh2007}. The modeled beam
width is directly related to the extent of the employed interpolation
or sampling kernel, and an approximate modeling of the beam width
(neglecting e.g.\ divergence) has been found to be sufficient in
practice \cite{HofmannKachelriess2014}. Joseph's 2D projector in
particular straight forwardly performs linear interpolation among
the nearest neighbors of each sampling point, which may as well be
interpreted as an approximation to normalized radial basis function
interpolation within a tightly limited radius. The modeled beam width
thus approximately corresponds to the voxel raster spacing. It has
been extended to 3D in the past by several authors to e.g.\ trace
X-rays through parallel stacks of textured planes \cite{XuMueller2005,XuMueller2006}
or for list mode reconstructions in positron emission tomography \cite{Schretter2006},
and is also found in recent reconstruction toolkits \cite{Rit2013,AarleSijbers2016}.
More elaborate calculations of line integrals over multilinearily
interpolated grids \cite{Koehler2000} have not been found to provide
practical benefits \cite{Turbell2001PhD}, and neither has the distance
driven method \cite{HahnNoo2016}. In their reviews of the field,
Pan et al.\ and Nuyts et al.\ similarly conclude that the particular
interpolation method is generally secondary as compared to an adequate
resolution of the voxel grid with respect to the features it is supposed
to represent \cite{Pan2009,Nuyts2013}. Eventually, an increasing
consensus can be identified supporting both the eligibility and the
sufficiency of basic interpolated sampling approaches. 

Considering computational efficiency again, it is preferable to keep
both the interpolation kernel size and the grid resolution to a necessary
minimum. Various strategies have been used to push that optimum beyond
localized kernels by using grids with adaptive resolution \cite{IhrkeMagnor2006,LeeuwenBatenburg2014}
and non-cartesian layout \cite{SitekGullberg2006,Scheins2015} or
even unstructured point clouds \cite{Gregson2012}. With regard to
cache efficiency of the given hardware, the layout of the volume image
data in memory may be better arranged with regard to expected access
patterns using e.g.\ techniques such as Z-ordering or blocking \cite{Beyer2014}.
Similarly, the algorithm design may be optimized to better follow
a given memory layout \cite{deManBasu2004,ThompsonLionheart2014}.
A central drawback of these more elaborate approaches to the reduction
and optimization of memory accesses is the increased algorithmic complexity,
limiting the net performance gain. In the case of very small grids
(typically less than $10^{6}$ voxels) and particularly when high
degrees of symmetry can be exploited, precalculation and explicit
storage of the sparse system matrix describing the projection process
can be an option as well, as addressed e.g.\ by \cite{Scheins2015}.
Finally, when simultaneously calculating large amounts of X-ray projections
of the same volume, divide and conquer approaches allow to systematically
reduce the amount of total memory accesses by exploiting spatial overlaps
of rays from close by viewing angles \cite{BrokishBresler2010}. For
parallel beam geometries, this can also be achieved by evaluating
projections in Fourier space \cite{MatejFessler2004}, based on the
Fourier slice theorem. In the present work, the efficient calculation
of individual projections within read-and-write memory will be addressed,
whereby efficiency will be drawn from simplicity as opposed to managing
irregular grids or memory layouts.

Starting with SGI graphics workstations in the 1990s, researchers
have further been utilizing the processing power of dedicated graphics
processors (GPUs) in order to speed up CT reconstruction. Reviews
on the usage of GPUs in tomography have been given e.g.\ by Mueller,
Pratx, Desprès and co-authors  \cite{Mueller2007,PratxXing2011,Despres2017}.

The aim of the present work is to demonstrate a ray driven projection
algorithm whose memory efficiency is implicit in its coherent sampling
pattern among parallel threads, and which is further formulated in
a computationally lean way. It thereby allows to perfectly utilize
the specific capabilities of modern general purpose graphics processors,
finally resolving the common conflict between sampling quality and
processing speed. 

\section{Methods\label{subsec:fwp-gjp}}

\subsection{Driving axis aligned grid traversal}

\begin{figure*}
\centering{}%
\begin{minipage}[b][1\totalheight][t]{0.43\textwidth}%
\begin{center}
\includegraphics[width=1\textwidth]{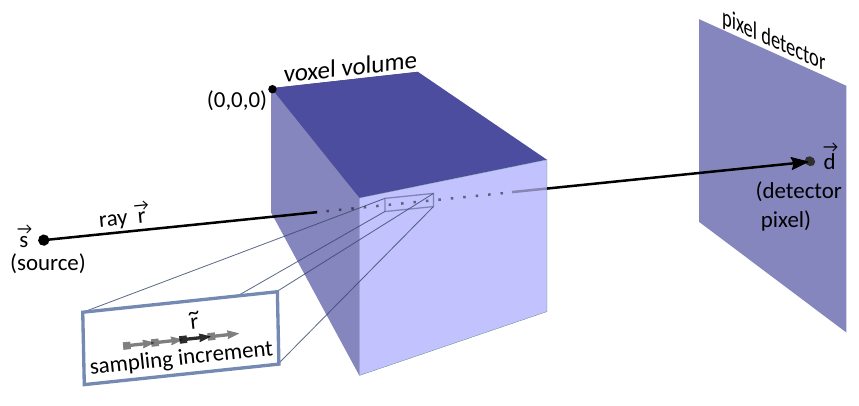}\\
~\\
\includegraphics[width=0.9\columnwidth]{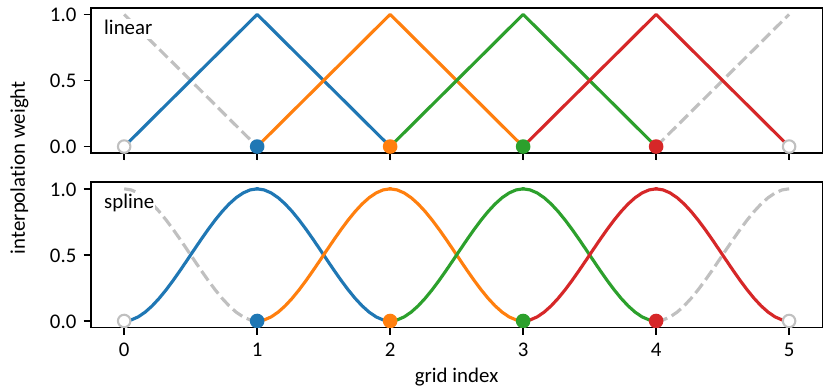}\\
\includegraphics[width=0.5\columnwidth]{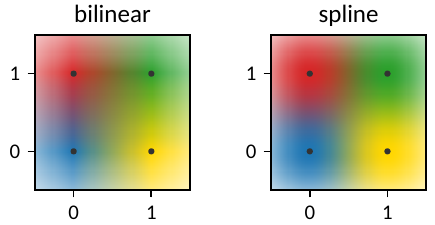}
\par\end{center}%
\end{minipage}$\;$\raisebox{0.5cm}{\includegraphics[width=0.55\textwidth]{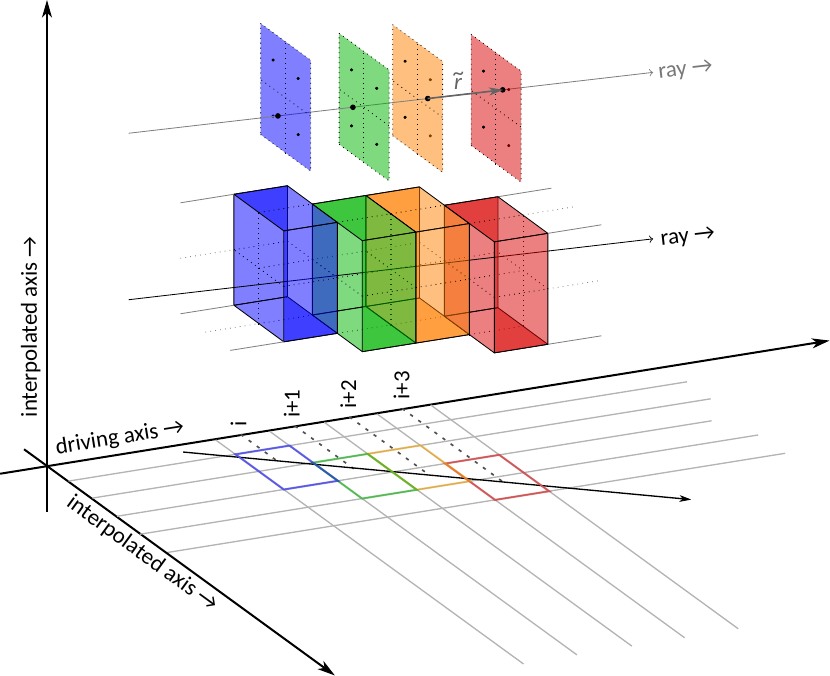}}\caption{\label{fig:gridaligned-raycasting}Ray casting through a volume along
a line defined by two points $\vec{s}$ and $\vec{d}$ (upper left).
On the right, driving-axis aligned sampling is illustrated on the
grid scale. Integer steps along the driving axis imply increments
of $\tilde{r}=\vec{r}/r_{m}$ along the actual path, with $\protect\smash{1\protect\leq\bigl\Vert\tilde{r}\bigr\Vert\protect\leq\sqrt{3}}$.
Due to the alignment with one designated axis, interpolation is required
only within 4-voxel blocks extending along the remaining axes (four
of such blocks are shown in different colors). Above, the corresponding
sampling points $\vec{p}^{\,(i)}$ are illustrated within their sampling
planes described by the respective nearest neighbour locations $\vec{v}^{(i,1-4)}$.
On the center left, the one dimensional interpolation kernels resulting
from Equations \ref{eq:interp-linear-1d} and \ref{eq:interp-spline-1d}
are illustrated. Each marker represents a point on the grid, and the
color-matched kernels represent the respective interpolation weights
as a function of fractional position between grid points. The resulting
two dimensional interpolation scheme is demonstrated below. }
\end{figure*}

The basic sampling concept is illustrated in Figure~\ref{fig:gridaligned-raycasting}.
A ray emanating from a source at $\vec{s}$ traverses a voxel volume
and hits a detector pixel at $\vec{d}$. Along its intersection with
the volume, the latter will be sampled in steps of $\tilde{r}$, which
will be concretized in the following. While the resulting scheme is
equivalent to general driving axis based methods, the present vector
representation allows for a unified treatment of all cases, such that
the ``driving axis'', which normally distinguishes different code
branches, now only determines the orientation of sampling planes within
a branchless sampling loop.

Given the positions of source $\vec{s}$ and detector pixel $\vec{d}$
relative to the volume origin, the integration path is characterized
by the set of points $\vec{p}$
\begin{align}
\vec{p} & =\vec{s}+l\vec{r}\\
\shortintertext{\text{with}}\vec{r} & =\vec{d}-\vec{s}
\end{align}
and $l\in\mathbb{R}$ being the free parameter. The driving axis $m$
is then identified by the largest component of $\vec{r}$:
\begin{equation}
m=\underset{i}{\mathrm{argmax}}(|r_{i}|)\:.
\end{equation}
The increment vector $\tilde{r}$ between successive sampling points
will be chosen such that the resulting sampling points remain aligned
with the driving axis, which holds for
\begin{equation}
\tilde{r}=\frac{\vec{r}}{r_{m}}\:.
\end{equation}
Assuming that grid coordinates correspond here to non-negative memory
indices, the first possible sampling point is defined by the intersection
of the ray with a plane through the origin and perpendicular to the
driving axis $m$, i.e.,
\begin{equation}
\begin{aligned}[][\vec{s}+o\,\tilde{r}]_{m} & \overset{!}{=}0\\
\Rightarrow o & =-s_{m}
\end{aligned}
\end{equation}
where $o$ is the distance between source and first sampling plane
in units of the sampling increment $\bigl\Vert\tilde{r}\bigr\Vert$.
$o$ will thus be termed ``sampling offset''. The volume can now
be sampled at points $\vec{p}^{\,(i)}$ along the defined path in
unit steps of axis $m$ by evaluating
\begin{equation}
\vec{p}^{\,(i)}=(\vec{s}+o\,\tilde{r})+i\,\tilde{r}
\end{equation}
for integer $i\in[0,i_{\max}${]}, where $i_{\max}$ is defined by
the extent of the voxel grid along axis $m$. These sampling points
can readily be used for linear interpolated sampling, as is e.g.\
directly provided by texture memory of modern GPUs. 

\subsection{Interpolated sampling}

When sampling from (GPU) main memory, the 4-neighborhood $\{\vec{v}^{(i,1)},\vec{v}^{(i,2)},\vec{v}^{(i,3)},\vec{v}^{(i,4)}\}$
of integer valued grid coordinates around each sampling point $\vec{p}^{\,(i)}$
needs to be explicitly enumerated. The driving axis component $p_{m}^{(i)}$
is, by construction of the sampling increment $\tilde{r}$ and offset
$o$, guaranteed to be integer for all integer $i$. The remaining
non-integer components necessarily lie between two integer ones along
their respective coordinate axes. For each sampling point $\vec{p}^{\,(i)}$,
the set of four neighboring voxels can thus be determined by regarding
all combinations of floor and ceiling values of these non-integer
components (with $\bigl\lfloor\,\bigr\rfloor$ and $\bigl\lceil\,\bigr\rceil$
being the floor and ceiling operators respectively):
\begin{equation}
\begin{aligned}\vec{v}^{(i,1)} & =\bigl\lfloor p_{1}^{(i)}\bigr\rfloor,\bigl\lfloor p_{2}^{(i)}\bigr\rfloor,\bigl\lfloor p_{3}^{(i)}\bigr\rfloor\\
\vec{v}^{(i,2)} & =\bigl\lfloor p_{1}^{(i)}\bigr\rfloor,\bigl\lceil p_{2}^{(i)}\bigr\rceil,\bigl\lceil p_{3}^{(i)}\bigr\rceil\\
\vec{v}^{(i,3)} & =\bigl\lceil p_{1}^{(i)}\bigr\rceil,\bigl\lfloor p_{2}^{(i)}\bigr\rfloor,\bigl\lceil p_{3}^{(i)}\bigr\rceil\\
\vec{v}^{(i,4)} & =\bigl\lceil p_{1}^{(i)}\bigr\rceil,\bigl\lceil p_{2}^{(i)}\bigr\rceil,\bigl\lfloor p_{3}^{(i)}\bigr\rfloor\:,
\end{aligned}
\label{eq:interp-nn}
\end{equation}
exploiting that 
\begin{equation}
\bigl\lfloor p_{m}^{(i)}\bigr\rfloor=\bigl\lceil p_{m}^{(i)}\bigr\rceil=p_{m}^{(i)}\:.
\end{equation}
Independent of $m\in\{1,2,3\}$, the vectors $\vec{v}^{(i,1-4)}$
define a group of four voxels in a plane perpendicular to the driving
axis. Illustrations of the planes spanned by these nearest neighbor
voxels around sampling points $\vec{p}^{\,(i)}$ are given in Figure~\ref{fig:gridaligned-raycasting}.
Special cases arise when either of the non-$m$ components of $\vec{p}^{\,(i)}$
happen to be also integer, which leads to redundant vectors among
$\vec{v}^{(i,1-4)}$. Given the final objective of interpolation,
these cases will be accounted for by adequate choice of the respective
interpolation weights.

Interpolation will be based on scalar distance weights 
\begin{equation}
\begin{aligned}w(d) & ;\quad d\in[0,1]\\
\text{with}\quad w(1-d) & =1-w(d)
\end{aligned}
\end{equation}
with respect to the component wise distances of the contributing grid
points next to a sampling point:
\begin{align}
d_{k}^{(\mathrm{fl})} & =p_{k}^{(i)}-\bigl\lfloor p_{k}^{(i)}\bigr\rfloor\,;\qquad d_{k}^{(\mathrm{cl})}=1-d_{k}^{(\mathrm{fl})}\:,
\end{align}
where the superscripts $(\mathrm{fl})$ and $(\mathrm{cl})$ indicate
distances to the integer grid indices below and above the components
$p_{k}^{(i)}$ of $\vec{p}^{\,(i)}$. The definition of $d_{k}^{(\mathrm{cl})}$
as complement to $d_{k}^{(\mathrm{fl})}$ guarantees correct interpolation
weights also in the special case of integer components $p_{k}^{(i)}$,
where floor and ceiling values coincide.  When explicitly defining
\begin{equation}
\begin{aligned}w_{\mathrm{fl},k} & =\mathrlap{w(d_{k}^{(\mathrm{fl})})}\hphantom{1-w_{\mathrm{fl},k}\quad}\text{for }k\neq m\\
w_{\mathrm{cl},k} & =1-w_{\mathrm{fl},k}\quad\text{for }k\neq m\\
w_{\mathrm{fl},m} & =w_{\mathrm{cl},m}=1
\end{aligned}
\end{equation}
the interpolation weights $w^{(i,1-4)}$ for the respective voxels
$\vec{v}^{(i,1-4)}$ can be conveniently represented as:
\begin{equation}
\begin{aligned}w^{(i,1)} & =w_{\mathrm{fl},1}^{(i)}\cdot w_{\mathrm{fl},2}^{(i)}\cdot w_{\mathrm{fl},3}^{(i)}\\
w^{(i,2)} & =w_{\mathrm{fl},1}^{(i)}\cdot w_{\mathrm{cl},2}^{(i)}\cdot w_{\mathrm{cl},3}^{(i)}\\
w^{(i,3)} & =w_{\mathrm{cl},1}^{(i)}\cdot w_{\mathrm{fl},2}^{(i)}\cdot w_{\mathrm{cl},3}^{(i)}\\
w^{(i,4)} & =w_{\mathrm{cl},1}^{(i)}\cdot w_{\mathrm{cl},2}^{(i)}\cdot w_{\mathrm{fl},3}^{(i)}
\end{aligned}
\label{eq:interp-weights}
\end{equation}
without requiring further explicit consideration of the particular
driving axis $m$.

Two specific weighting functions shall be considered: 
\begin{align}
w_{\mathrm{lin}}(d) & =1-d\label{eq:interp-linear-1d}\\
w_{\mathrm{spl}}(d) & =1-3d^{2}+2d^{3},\label{eq:interp-spline-1d}
\end{align}
with $w_{\mathrm{lin}}$ reproducing classic multilinear interpolation
and $w_{\mathrm{spl}}$ being a smooth spline function in the style
of a smooth cosine window that ensures differentiability also at grid
points, i.e., when $d=0$ or $d=1$.  The practical consequences
of the different interpolations schemes are later addressed in Sections~\ref{subsec:fwp-proj-quality}--\ref{subsec:fwp-recon-quality}
and Figures~\ref{fig:projection-errors}--\ref{fig:reconstruction-results}
therein.

Algorithm~\ref{alg:branchless-gjp} combines the above considerations
on volume traversal, implicit identification of sampling planes, and
interpolation among the respective nearest neighbors into a single
sampling loop. As can be verified by explicitly assuming different
driving axes, the weighted sampling performed in lines \ref{lst:gjp-line-ndinterp-start}--\ref{lst:gjp-line-ndinterp-end}
always correspond to a 2D-interpolation among the nearest neighbors
of the respective sampling point within a plane perpendicular to the
axis $m$.

\begin{algorithm}[H]
\algvspace
\begin{algorithmic}[1]
\small{
\State $\mathbf{assuming:}$ $\vec{s},\vec{d}$ defined in units of voxel grid indices
\State $\vec{r} \gets \vec{d}-\vec{s}$\Comment integration path orientation
\State $m \gets \mathrm{argmax}_{i}(|r_i|)$\Comment $m$: major (driving) axis
\State $\tilde{r} \gets \vec{r}/r_m$\Comment $\tilde{r}$: sampling increment vector
\State $o \gets -s_m$ \Comment $o$: sampling offset, $\vec{s}$: ray source point
\State $i_{\max} \gets$ volumeDimensions[$m$]\Comment number of sampling points
\State $a \gets 0$\Comment $a$: accumulator variable
\For{$i=0\,..\,i_{\max}-1$}\Comment iterate over sampling points\label{lst:gjp-line-sloop-start}
  \State $\vec{p} \gets \vec{s} + (o + i)\cdot\tilde{r}$\Comment $\vec{p}$: current sampling point
  \If {$\vec{p}$ \textbf{is in} volume} 
	  \State $\mathrlap{\vec{p}_\mathrm{fl}}\hphantom{\vec{p}_\mathrm{cl}} \gets \mathrm{floor}(\vec{p})$\Comment find \makebox[\widthof{upper}][c]{lower} voxel grid indices\label{lst:gjp-line-sampling-start}
	  \State $\vec{p}_\mathrm{cl}  \gets  \hphantom{\mathrm{floor}}\mathllap{\mathrm{ceil}}(\vec{p})$\Comment find upper voxel grid indices
 \State{assert: $p_{\mathrm{fl},m} = p_{\mathrm{cl},m} = p_m$}
	  \State $\mathrlap{\vec{w}_\mathrm{fl}}\hphantom{\vec{w}_\mathrm{cl}} \gets w(\vec{p}-\vec{p}_\mathrm{fl})$\Comment interpolation weights (Eqs.\ \ref{eq:interp-linear-1d}, \ref{eq:interp-spline-1d}\label{lst:gjp-line-interp-func})
       \State $\vec{w}_\mathrm{cl} \gets \vec{1}-\vec{w}_\mathrm{fl}$ \label{lst:gjp-line-compweights}\Comment complementary weights

	  \State w$_{\mathrm{cl},m} \gets w_{\mathrm{fl},m} \gets 1$ \label{lst:gjp-line-w-idx-acs}\Comment special case: driving axis\label{lst:gjp-maxis-weights}
	  
	  \State $a \gets a+{}$volume$[\,\mathrlap{p_{\mathrm{fl},1}}\hphantom{p_{\mathrm{cl},1}}$, $\mathrlap{p_{\mathrm{fl},2}}\hphantom{p_{\mathrm{cl},2}}$, $\mathrlap{p_{\mathrm{fl},3}}\hphantom{p_{\mathrm{cl},3}}]$
                            $\cdot\left\Vert\tilde{r}\right\Vert\cdot w_{\mathrm{fl},1} \cdot w_{\mathrm{fl},2} \cdot w_{\mathrm{fl},3}$ \label{lst:gjp-line-ndinterp-start}
	  \State $a \gets a+{}$volume$[\,\mathrlap{p_{\mathrm{fl},1}}\hphantom{p_{\mathrm{cl},1}}$, $p_{\mathrm{cl},2}$, $p_{\mathrm{cl},3}]$
                            $\cdot\left\Vert\tilde{r}\right\Vert\cdot w_{\mathrm{fl},1} \cdot w_{\mathrm{cl},2} \cdot w_{\mathrm{cl},3}$
	  \State $a \gets a+{}$volume$[\,p_{\mathrm{cl},1}$, $\mathrlap{p_{\mathrm{fl},2}}\hphantom{p_{\mathrm{cl},2}}$, $p_{\mathrm{cl},3}]$
                            $\cdot\left\Vert\tilde{r}\right\Vert\cdot w_{\mathrm{cl},1} \cdot w_{\mathrm{fl},2} \cdot w_{\mathrm{cl},3}$
	  \State $a \gets a+{}$volume$[\,p_{\mathrm{cl},1}$, $p_{\mathrm{cl},2}$, $\mathrlap{p_{\mathrm{fl},3}}\hphantom{p_{\mathrm{cl},3}}]$
                            $\cdot\left\Vert\tilde{r}\right\Vert\cdot w_{\mathrm{cl},1} \cdot w_{\mathrm{cl},2} \cdot w_{\mathrm{fl},3}$ \label{lst:gjp-line-ndinterp-end}\label{lst:gjp-line-sampling-end}
  \EndIf
\EndFor .\label{lst:gjp-line-sloop-end}
} 
\end{algorithmic}
\algvspace 

\caption{\label{alg:branchless-gjp}Branchless ray casting through a cuboid
voxel volume with axis aligned sampling and first order interpolation
among four nearest neighbors. Scaling by $\bigl\Vert\tilde{r}\bigr\Vert$
accounts for the varying sampling distances depending on the orientation
of the ray $\vec{r}$. 2D X-ray projection images of a volume are
obtained by parallel evaluation for multiple rays defined by a focal
point $\vec{s}$ and multiple detector pixel locations $\vec{d}$.
Highest GPU cache efficiency is achieved when the driving axis $m$
does not coincide with the fastest changing index of the memory layout.}
\end{algorithm}

\section{Results\label{sec:fwp-benchmarks}}

\subsection{Quality of projection images\label{subsec:fwp-proj-quality}}

\begin{figure}
\begin{centering}
\includegraphics[width=0.95\columnwidth]{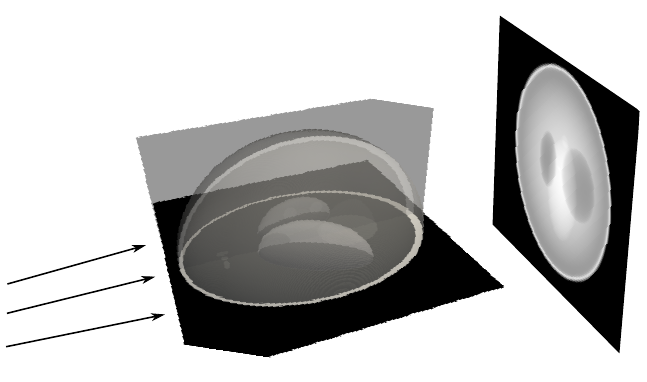}\\
~
\par\end{centering}
\begin{centering}
\includegraphics[width=1\columnwidth]{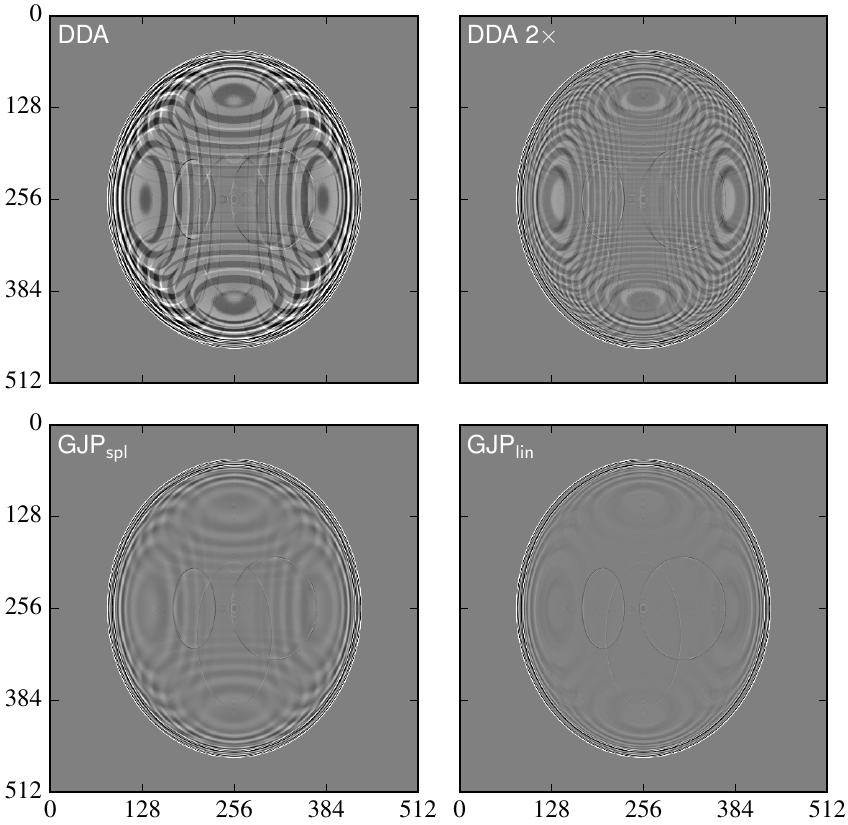}\\
~
\par\end{centering}
\begin{centering}
\includegraphics[width=0.9\columnwidth]{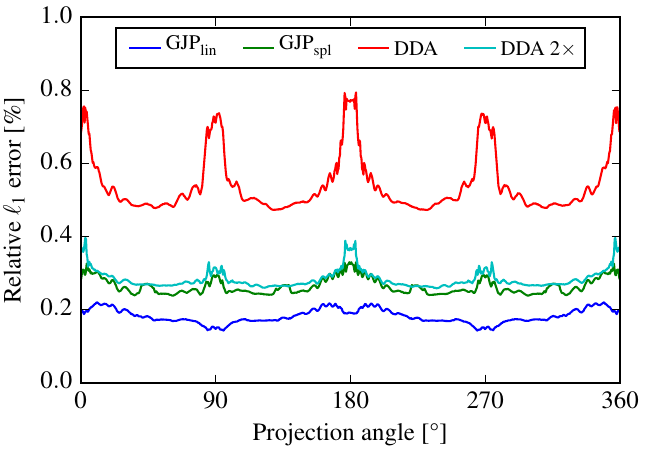}
\par\end{centering}
\caption{\label{fig:projection-errors}Approximation errors of different projection
algorithms for a 10° conebeam geometry. Numeric projections of the
rasterized Shepp Logan phantom (on a $512^{3}$ grid) onto a $512^{2}$
detector are compared to corresponding reference projections obtained
by analytic integration of the ellipsoids defining the phantom. At
the top, an illustration of the Shepp Logan phantom and the applied
projection geometry is given. Below, example difference images for
a frontal view are shown. On the bottom, the $\ell_{1}$ norm of these
residuals, normalized to the $\ell_{1}$ norm of the analytic reference
projection, is plotted for all projection angles of a circular scanning
trajectory within the axial plane.}

\end{figure}
The performance with respect to adequate modeling of of ray-volume
intersections is demonstrated on cone beam projections of a classic
three dimensional Shepp Logan phantom based on the definition reproduced
in \cite{Schabel2006}. The phantom is described by a sum of ellipsoids,
which can on the one hand be easily rasterized at any desired resolution
and on the other hand allows the direct calculation of projection
images by analytical evaluation of line integrals over the defining
ellipsoids. A ground truth is thus available for comparison with respective
numeric projections calculated from the rasterized version. In order
to also adequately account for the extent and integrating nature of
detector pixels, the reference projections are evaluated as an average
over 64 line integrals between the focal spot and regular arrays of
$8\times8$ points within each detector pixel. Analogously, oversampling
is applied also in the rasterization process of the phantom: It is
rasterized on a regular grid of $512^{3}$ voxels, whereby each voxel
value is determined as an average over $5\times5\times5$ regularly
distributed samples of the function defining the phantom.

Following typical experimental conditions, a cone angle of 10° is
modeled (i.e., the focal distance is about $5.7$ times the detector
width), projecting the volume onto a square detector of $512^{2}$
pixels. In total, 803 ($\approx\frac{\pi}{2}512$) projection images
from different orientations covering a full circle are computed, whereby
the chosen number of projections corresponds to a common recommendation
with regard to analytic tomographic reconstruction (cf.\ \cite{Bugug2011}).

Figure \ref{fig:projection-errors} shows residual projection errors
observed for various numeric projection approaches. Although the general
occurrence of such residuals is generally expected due to the inherently
approximative nature of discrete volume represenations, the adequacy
of a projection model may reasonably be measured by its ability to
keep such residuals minimal. Siddon's pencil beam projection model,
realized using the DDA algorithm, exhibits most artifacts, particularly
in cases where rays run roughly parallel to grid axes. In these situations
the model of pencil beams intersecting box-shaped voxels is equivalent
to nearest neighbor sampling. When oversampling the DDA by a factor
of two in each dimension in order to approximate a finite beam extent,
i.e. tracing and averaging over four rays per detector pixel, the
resulting projection residuals become comparable to those of the non-oversampled
GJP$_{\mathrm{spl}}$ algorithm using spline interpolation, although
the latter further shows a considerable reduction of high frequency
artifacts. The best results are, despite the kinked interpolation
kernel, achieved by the liner interpolating GJP$_{\mathrm{lin}}$
algorithm.

\subsection{Quality of iterative tomographic reconstructions\label{subsec:fwp-recon-quality}}

\begin{figure}
\begin{centering}
\includegraphics[width=0.9\columnwidth]{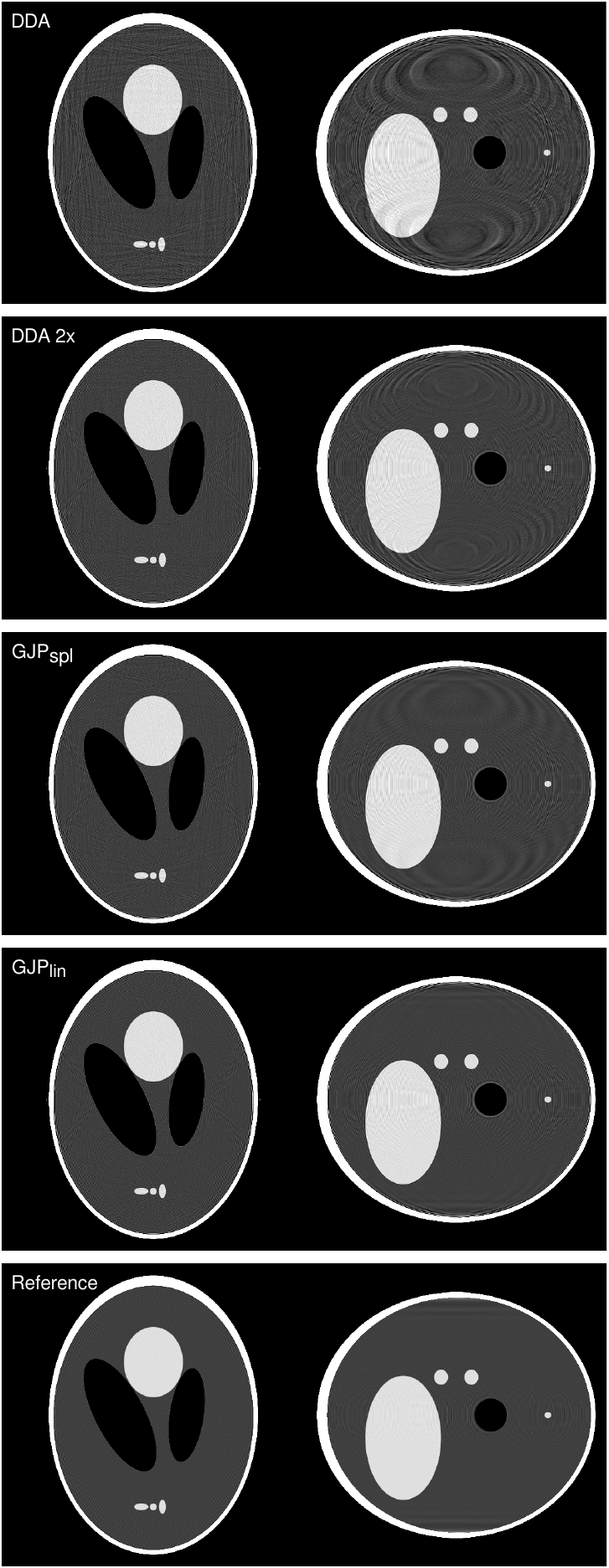}
\par\end{centering}
\caption{\label{fig:reconstruction-results}Axial (left) and sagittal (right)
central slices of iterative SART reconstructions (10 iterations) on
a discrete voxel grid from analytic projections of a 3D Shepp Logan
phantom using different numeric projection methods within the iterative
process. The chosen grayscale window shows a range of {[}0.16,0.32{]}
out of the maximal range of {[}0,1{]}. Limitations of the discrete
forward models (cf. Fig.~\ref{fig:projection-errors}) manifest themselves
in the final reconstruction result. At the bottom, a reconstruction
from perfect input data (synthesized with GJP$_{\mathrm{lin}}$ from
the rasterized Shepp Logan phantom) is shown for reference, demonstrating
the artifacts expected from the reconstruction process itself. Wavy
features at the sagittal top and bottom represent cone beam artifacts.}

\end{figure}
As iterative reconstruction techniques such as SART (\cite{AndersenKak1984},
``Simultaneous Algebraic Reconstruction Technique'') subsequently
enforce consistency of the reconstructed volume with each experimentally
observed projection image based on a given forward model (formally
represented by a matrix $\boldsymbol{A}$, applied to a vector of
volume elements $\boldsymbol{f}$, yielding a set of line integrals
$\boldsymbol{g}$), inaccuracies of the respective discrete forward
projectors will directly translate to artifacts in the reconstruction
result. In contrast to typical artifacts arising when reconstructing
from an under-determined system of equations, e.g., when reconstructing
from too few projections, deficiencies of the projection model defining
the system matrix $\boldsymbol{A}$ inherently do not lie in its null
space, and can therefore not, without loss of resolution, be compensated
by typical regularization approaches that are otherwise used to suppress
artifacts emerging \emph{in the null space} of $\boldsymbol{A}$,
i.e., in image domains that are not affected by $\boldsymbol{A}$
and its defining projection model.

In order to illustrate the practical consequences, multiple SART reconstructions
are compared using different projection algorithms. In general, such
experiments typically suffer either from unrelated artifacts when
working with actual experimental data, or from the ``inverse crime''
that is often committed when synthesizing experimental data based
on the same algorithms that are also used in the subsequent reconstruction
procedure. Both issues can however be avoided in the case of the Shepp-Logan
phantom due to the possibility to analytically calculate its projections
without prior rasterization, as has been done already for the previous
benchmark. 

Figure~\ref{fig:reconstruction-results} shows central axial and
sagittal slices of respective SART reconstructions on a $512^{3}$
voxel grid of the Shepp-Logan phantom from analytically calculated
projections as described previously. As volume rasterization is here
only introduced with the discrete imaging model fundamental to iterative
reconstruction techniques, the observed reconstruction artifacts can
be largely attributed to the employed discrete projection method.
While other parameters such as iteration count or the interpolation
scheme of the voxel based backprojector can also be argued to affect
the reconstruction outcome, it should nevertheless be without doubt
that these, in contrast to the forward model, do not actually explicitly
define the properties of the solution. The bottom panel of Fig.~\ref{fig:reconstruction-results}
illustrates the isolated effect of the present reconstruction procedure
by using ideal input data synthesized with the same forward model
(GJP$_{\mathrm{lin}}$) also used for iterative reconstruction. In
contrast to artifacts arising from the forward model itself, pure
reconstruction artifacts can generally be suppressed by regularization
techniques, which have here explicitly not been applied to avoid ambiguities
in the interpretation of the presented results. 

In accordance with the previously found projection errors shown in
Fig.~\ref{fig:projection-errors}, the reconstruction quality is
found to be worst for the non-oversampling DDA, comparable for 2-fold
oversampled DDA and GJP$_{\mathrm{spl}}$ and best for GJP$_{\mathrm{lin}}$.
Although it is out of the scope of the present work to explicitly
demonstrate the effect of each SART parameter, the reader shall however
be assured that variations in iteration count, relaxation factor and
backprojection interpolation scheme have been confirmed to not fundamentally
change the relative performance of different forward models. This
is in accordance with the preceding reasoning attributing the differing
artifact patterns to differing discretization errors among the various
methods.

Finally, these results are further independent of additive noise,
which has here intentionally not been regarded. As SART is, despite
being iterative, a linear reconstruction technique, additive terms
to the projection images can generally be considered independently
and will, although adding to the reconstruction result, not fundamentally
alter it.

\subsection{Projection speed}

Run time performance is evaluated for projections of a cylindric
volume (as common for tomographic reconstructions) within a cubic
bounding box of $512^{3}$ voxels onto a $512^{2}$ pixel detector.
The sampling offset $o$ marking the first sampling point and the
total number of sampling points $i_{\max}$ are adapted to ray-cylinder
intersections as opposed to ray-boundingbox intersections to this
purpose. The performance of Algorithm~\ref{alg:branchless-gjp} is
benchmarked against the branchless DDA formulation given by \cite{Xiao2012}.
The volume is stored in 32bit floating point format in either main-
or texture memory of the graphics processing unit. For the case of
texture memory, also hardware provided interpolation is tested. As
typical for computed tomography setups, projections are performed
for a multitude of source and detector orientations over the full
angular range of $360\text{°}$ on a circular trajectory around the
volume center. The rotational axis is aligned parallel to the fastest
index of the memory layout (i.e., the last dimension in the case of
Fortran-style memory order, or the first dimension in the case of
C-style memory order). For each individual configuration of source
and detector, the run time is optimized over a wide range of possible
thread block or work group size parameters (CUDA and OpenCL terminology
respectively). This eliminates the potential influence of technicalities
introduced by the parallelization schemes of graphics processors.
Measured execution times further exhibit a variance of up to 10\%
when running the same code multiple times due to dynamic performance
adaptions related to temperature management. Reported are the fastest
measured times for each algorithm.

\begin{table*}
\centering{}%
\begin{tabular}{cccccccccccc}
\hline 
 & \multicolumn{2}{c}{DDA} &  & \multicolumn{2}{c}{DDA 2$\times$} &  & \multicolumn{2}{c}{GJP$_{\mathrm{lin}}$} &  & \multicolumn{2}{c}{GJP$_{\mathrm{hwlin}}$}\tabularnewline
 & GTX970 & GTX1080 &  & GTX970 & GTX1080 &  & GTX970 & GTX1080 &  & GTX970 & GTX1080\tabularnewline
\hline 
\hline 
\multirow{2}{*}{\begin{turn}{90}
Tex.~
\end{turn}} & 4.59$\,$ms & 3.46$\,$ms &  & 15.2$\,$ms & 6.74$\,$ms &  & 4.97$\,$ms & 2.40$\,$ms &  & 3.28$\,$ms & 2.23$\,$ms\tabularnewline
 & 118$\,$GB/s & 157$\,$GB/s &  & 143$\,$GB/s & 323$\,$GB/s &  & 334$\,$GB/s & 686$\,$GB/s &  & 502$\,$GB/s & 740$\,$GB/s\tabularnewline
\hline 
\multirow{2}{*}{\begin{turn}{90}
{\small{}RAM}$\;$
\end{turn}} & 4.39$\,$ms & 5.20$\,$ms &  & 15.7$\,$ms & 7.65$\,$ms &  & 5.69$\,$ms & 2.70$\,$ms &  & --- & ---\tabularnewline
 & 123$\,$GB/s & 104$\,$GB/s &  & 139$\,$GB/s & 285$\,$GB/s &  & 292$\,$GB/s & 609$\,$GB/s &  & --- & ---\tabularnewline
\hline 
\end{tabular}\caption{\label{tab:execution-times} Average projection speed in milliseconds
and memory access rates in gigabytes per second for 2D projections
of a cylindrical volume within a $512^{3}$ bounding box onto a $512^{2}$
detector in a 10° conebeam setup measured an Nvidia GTX 970 and GTX
1080 GPU.}
\end{table*}

Table~\ref{tab:execution-times} lists the so evaluated run times
as averages over 360 equidistant projection angles for two GPU models.
As a measure for GPU occupancy it further lists average memory access
rates based on the total runtime and the amount of accessed voxels
by each raytracing algorithm respectively. Although the latter is
not strictly known in the case of GJP$_{\mathrm{hwlin}}$ due to unknown
implementation details within the GPU, it is reasonably assumed to
be the same as for GJP$_{\mathrm{lin}}$. 

A number of interesting conclusions can be drawn from the observed
timings: First of all, the DDA algorithm can only benefit from newer
hardware (GTX 1080) in the oversampled case. Oversampling increases
the number of duplicate accesses to the same voxels by parallel threads
handling neighboring rays, wherefore it can be reasonably assumed
that the oversampled DDA better profits from memory caches. The additional
computational overhead associated with oversampling appears to be
a limiting factor on older hardware in contrast, where the overall
runtime increases almost linear with the amount of traced rays. This
assessment is consistent with the observation that the DDA algorithm
does not profit from optimized accesses to read only texture memory.
The GJP algorithm in contrast is able to outperform even the regular
DDA algorithm by a factor of up to 2, despite accessing about twice
as much memory on average. The driving axis aligned sampling scheme
of the GJP ensures that neighboring threads partially access the same
voxels in the course of interpolated sampling, thereby exploiting
memory caches even better than the oversampled DDA. The additional
speedup observed when simplifying the GJP algorithm even further (by
using the intrinsic interpolation capabilities of texture memory)
indicates that it operates close to the limits both of the computational
resources and the available memory bandwidth.

\section{Discussion and Conclusion}

The calculation of projections from discrete volumes is a core aspect
of iterative reconstruction techniques, both with respect to reconstruction
speed and quality. Although a remarkable variety of approaches to
the advanced modeling of ray-volume intersections has been presented
in previous literature, the demand for maximal parallelizability and
computational efficiency on modern graphics processors immediately
collapses the wide palette of choices to ray driven methods with strongly
confined sampling kernels. ``Ray driven'' thereby implies a sampling
loop iteratively traversing the voxel grid along defined paths (rays)
between focal point and detector pixels, and ``strongly confined''
implies the evaluation of only the immediate neighborhood around each
sampling point. For the traversal of regular grids, two methods can
be named: the digital differential analyzer (DDA) algorithm \cite{AmanatidesWoo1987},
traversing the grid in unevenly spaced steps from intersection to
intersection with any of the orthogonal grid planes, and methods traversing
the grid in equidistant steps aligned with a designated driving axis.
The former technique allows to precisely determine line-box intersection
lengths and corresponds to the much cited pencil-beam X-ray imaging
model given by Siddon \cite{Siddon1985}, while the latter technique
is typically combined with interpolated sampling and then corresponds
to the competing model proposed in the context of tomographic reconstruction
by Joseph \cite{Joseph1982}.

A branchless formulation of a Joseph type interpolating volume projection
algorithm has been derived here, with the particular benefit of being
extremely simple, which is a general prerequisite for maximal computational
efficiency. Driving axis aligned sampling ensures an optimal amount
of sampling points along each path in the sense that voxels are neither
skipped nor oversampled. The resulting synchronous progression of
parallel rays through the voxel grid thereby ensures high cache hit
rates without explicitly constraining the exact imaging geometry (as
opposed to e.g.\ the cache optimized Siddon's algorithm proposed
by \cite{ThompsonLionheart2014}, or the symmetry exploiting projection
model given by \cite{Scheins2015}). Interpolated sampling among the
remaining dimensions has been argued, besides being a practical necessity,
to be consistent with ideas on exact modeling of X-ray projections
based on normalized radial basis functions or projection footprints.
Approximate matching of the voxel grid spacing to the average density
of rays between focal point and detector array thereby ensures an
adequate modeling of beam width, implicitly reproducing the integrating
nature of detector pixels of finite extent without requiring far ranging
interpolation kernels or oversampled ray casting. In accordance with
assessments given in previous literature, higher order effects such
as cone beam related variations in beam extent can be safely neglected
in the modeling. (cf.~\ e.g.\ \cite{Nuyts2013,HofmannKachelriess2014})

The performed benchmarks compared a number of self-suggesting variants
of both ray casting algorithms with respect to artifacts and computational
efficiency, addressing the recurring questions of adequate beam shape
modeling and the role of the chosen interpolation kernel. The results
indicate that no tradeoff needs to be made between computational efficiency
and fitness for the purpose: the proposed simple and efficient branchless
3D Joseph projector employing linear interpolation is found to clearly
perform best both with regard to approximation of the ground truth
and with regard to efficiency, operating in the range of the theoretic
maximum capabilities of the employed hardware.

A recurring concern with regard to local interpolation exists in situations
where a sufficient matching of the voxel grid resolution to the detector
resolution is seemingly impossible. Such situations can e.g.\ arise
when attempting to combine isotropic volume sampling with highly asymmetric
detector pixels. It is in such cases obviously generally possible
to cast an adequate amount of rays per detector bin ensuring sufficient
coverage of the voxel grid, i.e., to adequately oversample the detector
image. Similarly, the voxel shape may be chosen non-square (in terms
of spatial units) to adequately match the detector properties, which
in units of grid indices does not alter the discussed algorithms.
Arguments questioning the representativeness of the Shepp-Logan phantom,
that has here been chosen for the sake of analytic integrability,
may be countered by noting that the observed artifacts arise at extended
material boundaries of moderate curvature, i.e., a situation that
is typical to CT applications.

The proposed formulation of a linearly interpolating Joseph-type projection
algorithm may eventually be considered a favorable choice in many
regards (simplicity of implementation, computational efficiency, and
fitness for the purpose) for typical CT reconstruction applications,
in particular as compared to the competing DDA algorithm, and further
considering that oversampling (i.e., increasing the density of traced
rays) generally remains an option.

\bibliographystyle{ieeetr}
\bibliography{}

\end{document}